# Observational Signatures of Coronal Heating
## in MHD Simulations Without Radiation or a Lower Atmosphere

James A. Klimchuk[1], Kalman J. Knizhnik[2], Vadim M. Uritsky[3]

## Abstract

It is extremely difficult to simulate the details of coronal heating and also make meaningful predictions of the emitted radiation. Thus, testing realistic models with observations is a major challenge. Observational signatures of coronal heating depend crucially on radiation, thermal conduction, and the exchange of mass and energy with the transition region and chromosphere below. Many magnetohydrodynamic simulation studies do not include these effects, opting instead to devote computational resources to the magnetic aspects of the problem. We have developed a simple method of accounting approximately for the missing effects. It is applied to the simulation output *post facto* and therefore may be a valuable tool for many studies. We have used it to predict the emission from a model corona that is driven by vortical boundary motions meant to represent photospheric convection. We find that individual magnetic strands experience short-term brightenings, both scattered throughout the computational volume and in localized clusters. The former may explain the diffuse component of the observed corona, while the latter may explain bright coronal loops. Several observed properties of loops are reproduced reasonably well: width, lifetime, and quasi-circular cross-section (aspect ratio not large). Our results lend support to the idea that loops are multi-stranded structures heated by "storms" of nanoflares.

## 1. Introduction

The processes that determine the thermal properties of the corona and its radiation spectrum involve an enormous range of spatial scales and physical couplings. Treating all these processes in a single numerical simulation is extremely challenging, if not currently impossible. The heating of the plasma is associated with magnetic reconnection at thin current sheets that are generated directly or indirectly by photospheric driving. There are on the order of 100,000 of these sheets in a single active region (Klimchuk 2015). Wave heating also requires very small spatial scales (Klimchuk 2006; Van Ballegooijen et al. 2011; Antolin et al. 2015). The response of the plasma to the heating involves radiation, field-aligned thermal conduction, flows, and a fundamental connection to the lower atmosphere. Both mass and energy are exchanged with the lower atmosphere, and any meaningful prediction of observations must account for this exchange, which requires an accurate treatment of the thin transition region. Neither current

---

[1] Heliophysics Science Division, NASA Goddard Space Flight Center, 8800 Greenbelt Rd, Greenbelt, MD 20771, USA
[2] Space Science Division, Naval Research Lab, 4555 Overlook Ave SW, Washington, DC 20375, USA
[3] Catholic University of America, 620 Michigan Ave NE, Washington, DC 20064, USA

sheets nor the transition region is stationary, so numerically resolving them in a realistic setting is extremely difficult even with nonuniform grids.

Because of these competing demands, many studies concentrate on limited aspects of the problem, foregoing others. For example, one-dimensional (1D) hydrodynamic simulations – often called loop models – treat the field-aligned physics and atmospheric coupling very well, but they apply to rigid flux tubes and require that coronal heating be specified as a model input. Many three-dimensional magnetohydrodynamic (3D MHD) simulations address the coronal heating question well, but they essentially ignore the plasma response by neglecting radiation, thermal conduction, and atmospheric coupling. Some ambitious MHD simulations include these missing effects, but the heating in those simulations comes from relatively passive Ohmic and viscous dissipation of broad current and velocity structures, rather than from explosive reconnection and associated shocks at a multitude of small current sheets. Whether the former is a reasonable proxy for the latter has yet to be established.

We report here on a method of estimating the time-dependent emission from MHD simulations that lack radiation, thermal conduction, and a lower atmosphere. The method is applied *post facto* to the simulation output. It is highly simplistic and no substitute for an eventual rigorous MHD treatment that includes the full physics, but it is a useful way to compare the simulations with observations to gain new insights or evaluate whether an idea is plausible. We refer to the method as a cooling model because plasma cooling is the essential missing ingredient in the MHD.

The model treats the evolution of the average pressure along a magnetic field line. It is based on three approximations. First, plasma heating is represented by increases in pressure in MHD simulations that exclude radiation. Second, plasma cooling – including the effects of radiation, thermal conduction, and atmospheric coupling – is assumed to produce an exponential decrease in pressure. Third, detected emission behaves similarly to pressure for observations made in temperature sensitive observing channels such as the 193 and 335 A channels of the Atmospheric Imaging Assembly (AIA) on the Solar Dynamics Observatory (SDO) (Lemen et al. 2012). We further explain and justify these approximations below.

We apply the model to our previously published simulation of coronal heating (Knizhnik et al. 2018). As discussed in that paper, we were able to extract valuable information about the statistics of impulsive heating events – nanoflares – but we were unable to say anything about their observational consequences. With the cooling model, we are now able to do so. As we discuss, our results suggest that the diffuse component of the corona is due to randomly scattered and seemingly uncorrelated nanoflares, while individual bright loops are due to clusters of events, or nanoflare "storms" (Klimchuk 2009).

## 2. Cooling Model

The response to a nanoflare of the plasma contained in a magnetic strand (elemental "loop") is well understood (Cargill 1994; Klimchuk 2006; Reale 2014). Temperature increases rapidly as the nanoflare occurs, leading to a greatly enhanced thermal conduction flux down the strand legs. This drives an upflow – known as chromospheric evaporation – that fills the strand and increases its density. Evaporation continues after the nanoflare ends, and the plasma cools from the conduction losses. These losses diminish as the temperature decreases, but radiation increases and eventually takes over as the dominant cooling mechanism. The strand then enters a phase where temperature and density decrease together as plasma drains and collects back onto the lower atmosphere.

Because thermal conduction and flows are so efficient at transferring energy and mass along the magnetic field, and because most strands are short compared to the gravitational scale height ($10^5$ km for $T = 2$ MK), pressure, temperature, and density all tend to be quite uniform along the strand. Only in the thin transition region at the base do temperature and density have large gradients. Pressure remains essentially constant through the transition region because the pressure scale height at the local temperature is everywhere much larger than the short temperature and density scale lengths. These well-known properties are the justification for the field-aligned hydrodynamics code Enthalpy-Based Thermal Evolution of Loops (EBTEL) (Klimchuk et al. 2008; Cargill et al. 2012). EBTEL computes the evolution of the spatially averaged coronal temperature, density, and pressure along a strand for a given time-dependent spatially averaged heating rate. Although the solutions are approximate, they agree well with exact solutions from 1D hydro codes that take several orders of magnitude more time to compute. Our cooling model, like EBTEL, treats the evolution of the average pressure in the strand.

The solid curve in Figure 1 shows the pressure from a 60,000 s EBTEL simulation in which a strand of $3 \times 10^9$ cm halflength is heated randomly by nanoflares of different energy. Each event has a triangular heating profile (symmetric rise and fall) with a total duration of 500 s. The energies were selected randomly from a power law energy distribution of slope -2.4 (Lopez Fuentes & Klimchuk 2016), and the delay between successive events is proportional to the energy of the first event. This corresponds to a scenario in which footpoint driving tangles and twists the magnetic strands until a critical misalignment angle is reached. The temporally averaged energy flux is $1.1 \times 10^7$ erg cm$^{-2}$ s$^{-1}$ and the median delay between successive events is 1180 s, both consistent with values inferred from active region observations (Klimchuk 2015, Klimchuk and Hinode Review Team 2019; Barnes, Bradshaw, & Viall 2021). As can be seen in Figure 1, there are times when nanoflares occur at high frequency, maintaining an approximately steady pressure, and times when nanoflares occur at low frequency, allowing substantial cooling between events.

Strands lose energy only by radiation. Thermal conduction, evaporation, and draining merely serve to transfer energy between the corona and lower atmosphere. Because the flows are subsonic – except perhaps during the earliest stages of especially energetic nanoflares – most of the plasma energy is thermal. Pressure, which is proportional to the thermal energy density, therefore increases as the nanoflare heating is occurring and decreases thereafter. Our cooling model assumes that it decreases exponentially in the absence of heating:

$P(t) = P_0 \, exp(-t/\tau)$, where the timescale $\tau$ is allowed to depend on pressure: $\tau \propto P^\alpha$. We do not have a rigorous explanation for this exponential form, given the complex evolution of described earlier, but we show below that it describes the pressure evolution remarkably well.

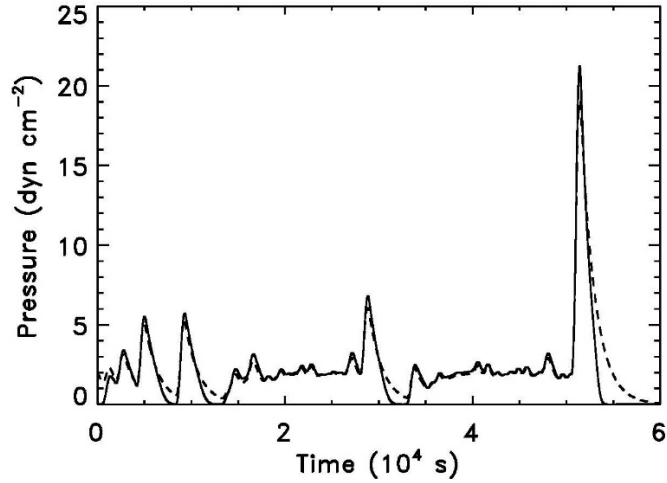

*Figure 1: Evolution of the strand-averaged coronal pressure for a strand that is heated randomly by nanoflares: solid – from the EBTEL output; dashed – from the cooling model based on the nanoflare heating input to EBTEL.*

The complete model, including both heating and cooling, updates the strand-averaged pressure in time step $\Delta t$ according to:

$$P_c(t + \Delta t) = P_c(t)\, exp\left[-\frac{\Delta t}{\tau_0}\left(\frac{P_0}{P_c(t)}\right)^\alpha\right] + \frac{2}{3}\, Q(t)\Delta t \ . \tag{1}$$

The subscript "c" indicates that this is the pressure of our cooling model, which we later distinguish from the pressure of the MHD simulation. The last term is the pressure increase associated with the energy input from heating, where $Q$ is the volumetric heating rate averaged along the strand (erg cm$^{-3}$ s$^{-1}$), and the ratio of specific heats is 5/3.

We have tried different values for the parameters of the cooling model and find that $\tau_0 = 1500$ s and $\alpha = 0$ ($P_0$ is irrelevant for this $\alpha$) give excellent agreement with the EBTEL simulation. The solid curve in Figure 1 shows the pressure from the EBTEL simulation, and the dashed curve shows the pressure predicted by Equation 1 using the same heating profile $Q(t)$ used with EBTEL. The two curves track very well.

We have not performed an exhaustive quantitative search for the best fit values of the parameters. Our goal is not to achieve a highly accurate model, but to account for the missing physics in MHD simulations to a degree that allows for a qualitatively prediction of the emission that would be produced if the missing physics were included. We note that $\alpha = 0$ (no dependence of the timescale on pressure) is similar to the extremely weak dependence derived for an impulsively heated loop by Cargill (1993).

It is important to understand that the timescale for pressure decrease, $\tau_0$, is different from the timescale for temperature decrease, often called the cooling time. Pressure decreases more slowly than temperature when evaporation is occurring, and it decreases more quickly when draining is occurring.

The primary purpose of our cooling model is to predict the emission that would be observed by instruments such as AIA on SDO. This detected emission depends on the square of density and a function of temperature that is different for each observing channel. The channels are designed to have temperature response functions that isolate a range of temperatures. This range can be narrow or broad depending on the channel.

Although the detected emission depends explicitly on temperature and density, we find that it behaves similarly to pressure when the heating is impulsive. There is not a direct connection with pressure, but the emission can be modeled in the manner of Equation 1. Specifically, the detected emissivity (erg cm$^{-3}$ s$^{-1}$) is given by

$$\varepsilon(t) = c\, P_c\big(t - t_{delay}\big) \, , \qquad (2)$$

where the pressure-to-emissivity conversion factor, $c$, and time delay, $t_{delay}$, are different for each channel. $P_c$ is from Equation 1, but with parameters that are channel dependent.

The reason for the time delay can be understood as follows. Consider a strand that is cooling after having been heated to high temperature by a nanoflare. Pressure peaks when the nanoflare ends, but the emission does not brighten until the plasma has cooled into the channel's range of temperature sensitivity. If the maximum nanoflare temperature is already in this range, there is still a delay in brightening because time is required for evaporation to increase the emission measure to a substantial value (density peaks after both temperature and pressure). The strand stays bright for a duration that depends on the width of temperature response function. This is captured by $\tau_0$, where broader channels have larger $\tau_0$.

Plasma also passes rapidly through a channel's temperature range as it is being heated by the nanoflare. However, densities tend to be very small at this time because evaporation has had little chance to operate, so the emission tends to be very faint. The cooling phase is much brighter and much longer lived.

We find that the emissivity detected in the 335 A channel of AIA can be reproduced with $c = 100$, $t_{delay} = 700$ s, $\alpha = 0$, and $\tau_0 = 1500$ s (the same $\alpha$ and $\tau_0$ used for pressure). This channel has maximum sensitivity near 3 MK and is quite broad (Viall & Klimchuk 2011). The solid curve in Figure 2 shows the emissivity as computed rigorously from the temperatures and densities of the EBTEL simulation in Figure 1 (coronal emission only; no transition region emission), and the dashed curve shows the emissivity predicted by Equations 1 and 2. The agreement is very acceptable. Discrepancies exist but are not crucial, as discussed below.

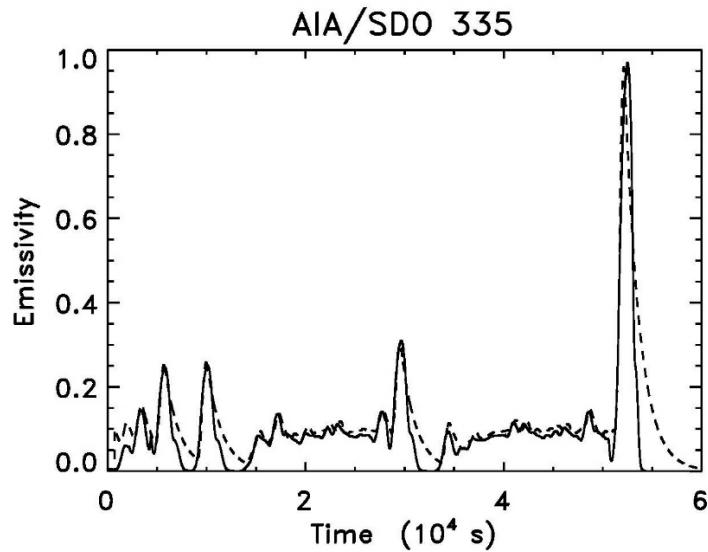

*Figure 2: solid – 335 A emissivity from the EBTEL temperature and density; dashed – predicted emissivity based on the cooling model and EBTEL pressure.*

Figure 3 shows the EBTEL and predicted emissivity in the 193 A channel. This channel is cooler and narrower than the 335 A channel, with maximum sensitivity near 1.5 MK. We therefore use a longer delay, $t_{delay}$ = 1500 s, and shorter duration, $\tau_0$ = 700 s. The agreement is further improved with $\alpha$ = 0.5 and $P_0$ = 2 dyn cm$^{-2}$. A larger conversion factor $c$ = 6000 accounts for the greater overall sensitivity of the channel. The model performs reasonably well, except for the last, most energetic event, where the duration of the brightening is greatly overestimated. We believe this can be explained by a catastrophic temperature collapse that sometimes occurs at the very end of cooling (Cargill & Bradshaw 2013; Reale & Landi 2012). This generally happens below 1 MK, but for especially energetic nanoflares, it can occur in the temperature range of the 193 A channel. This catastrophic cooling is not captured by our model. Fortunately, we are primarily interested in the collective emission from multiple strands, so accurately reproducing the detailed light curves of individual strands is not crucial.

An additional deficiency of the model is a tendency to overestimate the weakest emission in the 193 A channel. Again, this is not a significant problem because coronal observations are indicative of collective emission, which is dominated by the times when strands are bright.

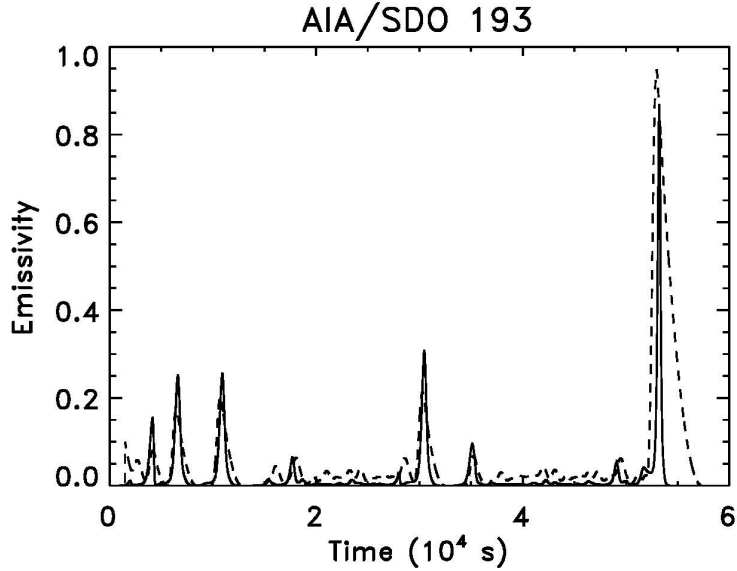

*Figure 3: solid – 193 A emissivity from the EBTEL temperature and density; dashed – predicted emissivity based on the cooling model and EBTEL pressure.*

These EBTEL simulations demonstrate that the cooling model performs well overall in predicting the emissivity in the 335 and 193 A channels whenever the heating takes the form of nanoflares having a duration of 500 s or less. Additional simulations are required to optimize the model parameters for other channels, different forms of heating, or strands of different length.

To apply the cooling model to our MHD simulation, we replace the heating term in Equation 1 with the change in the field-line-averaged pressure from the simulation, $P_{MHD}$:

$$P_c(t + \Delta t) = P_c(t)\, exp\left[-\frac{\Delta t}{\tau_0}\left(\frac{P_0}{P_c(t)}\right)^\alpha\right] + P_{MHD}(t + \Delta t) - P_{MHD}(t)\ . \qquad (3)$$

This is appropriate because there is no radiation in the simulation. Whereas $P_c$ rises and falls from heating and cooling, $P_{MHD}$ trends upward from the ongoing, uncompensated heating. We sometimes refer to $P_c$ as the cooled pressure and $P_{MHD}$ as the uncooled pressure.

The pressure in an MHD system is, however, affected by things other than direct heating. Unlike with EBTEL, which assumes a rigid flux tube, the magnetic field evolves. $P_{MHD}$ can increase or decrease from work done on or by the plasma as the strand volume changes. It can also increase or decrease because the two new strands that form from reconnection contain a mixture of plasma from the two original strands. EBTEL only includes direct heating and radiation. Fortunately, these tend to dominate. Compression heating is generally much weaker than direct heating from magnetic energy conversion[4], and we show below that the mixing effect is relatively small for the heating scenario we investigate here.

## 3. MHD Model

The MHD simulation was presented originally in Knizhnik et al. (2018). Beginning with a uniform "vertical" field in a box, we use the Adaptively Refined Magnetohydrodynamics Solver (ARMS; DeVore & Antiochos 2008) to solve the equations for mass, momentum, and energy conservation, and the magnetic induction equation:

$$\frac{\partial \rho}{\partial t} + \nabla \cdot \rho \boldsymbol{v} = 0 \ , \tag{4}$$

$$\frac{\partial \rho \boldsymbol{v}}{\partial t} + \nabla \cdot (\rho \boldsymbol{v}\boldsymbol{v}) = -\nabla P + \frac{1}{4\pi}(\nabla \times \boldsymbol{B}) \times \boldsymbol{B} \ , \tag{5}$$

$$\frac{\partial U}{\partial t} + \nabla \cdot \left\{ \left( U + P + \frac{B^2}{4\pi} \right) \boldsymbol{v} - \frac{\boldsymbol{B}(\boldsymbol{v}.\boldsymbol{B})}{4\pi} \right\} = 0 \ , \tag{6}$$

$$\frac{\partial \boldsymbol{B}}{\partial t} = \nabla \times (\boldsymbol{v} \times \boldsymbol{B}) \ , \tag{7}$$

where

$$U = \epsilon + K + W \tag{8}$$

is the total energy density – the sum of internal energy density

$$\epsilon = \frac{P}{\gamma - 1} \ , \tag{9}$$

kinetic energy density

$$K = \frac{\rho v^2}{2} \ , \tag{10}$$

and magnetic energy density

$$W = \frac{B^2}{8\pi} \ . \tag{11}$$

---

[4] Consider the release of magnetic energy in volume corresponding to a flux tube. Magnetic pressure decreases, and if the released energy is lost to radiation, the volume constricts to maintain pressure balance with the surroundings. (In our simulation, the magnetic energy is converted largely to thermal energy, with some kinetic, and there is minimal volume change.) The change in magnetic energy per unit length is $\delta E_m = (B^2/8\pi)\delta A$, where $\delta A$ is the decrease in cross sectional area. The work per unit length done by adiabatic compression of a similar volume of plasma is $P\delta A$. The ratio of the two is $\beta = 8\pi P/B^2$, which is very small in the corona.

In these equations, $\rho$ is mass density, $T$ is temperature, $P$ is pressure, $\gamma$ is the ratio of specific heats, $v$ is velocity, $B$ is magnetic field, and $t$ is time. There is no explicit resistivity or viscosity in the simulation, but ARMS has a minimal, though finite, numerical resistivity. This allows reconnection while conserving magnetic helicity extremely well, which is a key ingredient in accurately modeling high magnetic Reynolds number environments such as the solar corona (Knizhnik et al., 2015, 2017a, 2017b). Our chosen form for the energy equation rigorously conserves total energy. None disappears from numerical effects. Whatever energy is lost by the magnetic field is gained by the plasma.

The initial atmosphere, like the magnetic field, is uniform. There is no radiation or thermal conduction, nor is there a transition region or chromosphere. There is a photosphere only to the extent that the field is line tied at the top and bottom boundaries and cannot slip.

We drive the system with small-scale rotational motions at the top and bottom boundaries, which correspond to opposite polarity regions of the photosphere. The driving pattern consists of 199 closely packed vortex cells within a large hexagonal region. The rotation is equal and opposite at the two boundaries. The sense of rotation varies randomly from cell to cell, with a 3:1 preference for one direction over the other; thus, there is a net injection of helicity. The rotation rate ramps up and down such that one full turn of twist is imparted to the field over each cycle. The cycle then repeats, maintaining the same sense, for a total of 15 cycles. The relative phasing of the cells is random. The model and driving properties are discussed more fully in Knizhnik et al. (2018). Observations of small-scale rotational motions are discussed in Bonet et al. (2008), Wedemeyer-Bohm & Rouppe van der Voort (2009), and Wedemeyer-Bohm et al. (2012).

The equations are solved in dimensionless units on a 640x640x128 uniform numerical grid. We convert to physical units by specifying that the vertical height of the box (initial length of the magnetic strands) is $2 \times 10^4$ km, the initial magnetic field strength is 50 G, the plasma $\beta$ is 0.2, and the peak driver velocity is 1 km s$^{-1}$, which is 5% of the Alfven speed. These values imply an initial plasma pressure of 20 dyn cm$^{-2}$. This is about an order of magnitude larger than in the actual corona, but our cooling model quickly lowers this to realistic values. The vortices in the driver flow have a resulting diameter of 5000 km that is spanned by 32 grid cells. They are not meant to represent particular solar surface features, but are a convenient way of injecting energy into the field via small scale incompressible flow. We will explore other forms of driving in a future study. We note that the spatial resolution may not be adequate to properly treat the onset of reconnection in the current sheets that develop (Leake, Daldorff, & Klimchuk, 2020). How this impacts the results is an important question that applies to all MHD simulations of the corona and is one that we are actively pursuing. The total duration of the simulation is $2.03 \times 10^5$ s, or more than 2 days.

As the boundary flows are applied, the initially uniform field becomes progressively more stressed. Instabilities develop and reconnection occurs. A statistical steady state is established in which the Poynting flux of energy pumped into the field by the driving is balanced by the energy removed from the field by the reconnection. This energy remains in the system, mostly in the form of thermal energy, but with a small amount of kinetic energy. As discussed in Knizhnik et al. (2018), a complex and ever-changing web of current sheets is created. See Figure 4 of that paper. Although the pattern of driving remains simple and organized, the field line connectivity between the photospheric boundaries is not. The connections are far more complex than a collection of coherent twisted flux tubes, as would be the case if there were no reconnection.

The cooling model is applied to the average pressure along the field lines. We trace field lines upward and downward from a 400x400 grid on the midplane. This is nontrivial computationally, so we do so at intervals of 445 s, much longer than the MHD simulation time step, and shorter than the timescale in the cooling model, $\tau_0$. The time step of the cooling model, $\Delta t$, is also 445 s. The variation in pressure along individual field lines in the MHD simulation is typically of order 0.1 dyn cm$^{-2}$, though this is not important since the cooling model treats field line averages. The equilibration time for smoothing out pressure variations is the sound travel time and is comparable to $\Delta t$ midway into the simulation.

Figure 4 shows the evolution of the cooled pressure for a representative field line midway into the simulation. The interval covers 15,000 s. Figure 5 shows a 15,000 s interval from the EBTEL simulation (subset of the dashed curve in Fig. 1). The two curves have similar characteristics, such as the variety of peak amplitudes and separations, suggesting that the heating in the MHD simulation is not unlike that assumed for the EBTEL simulation, i.e., impulsive bursts that follow a power law energy distribution. This is not surprising. We showed in our original paper on this simulation that various proxies of heating have spatial and temporal distributions that obey power laws (Knizhnik et al. 2018; see also Knizhnik & Reep 2020 and Knizhnik et al. 2020).

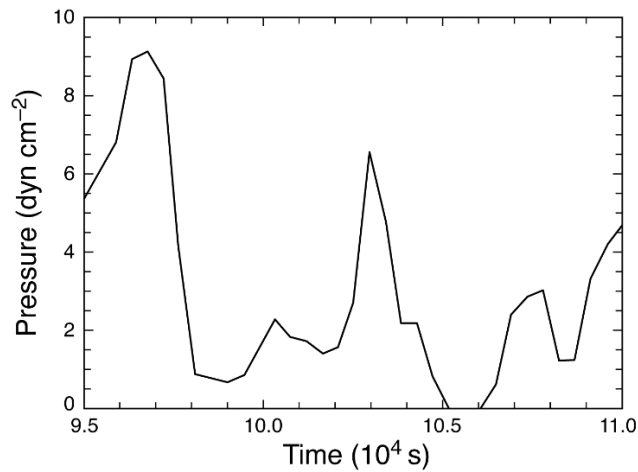

*Figure 4 : Pressure evolution over a 15,000 s interval at a representative grid point in the mid-plane of the combined MHD/Cooling simulation.*

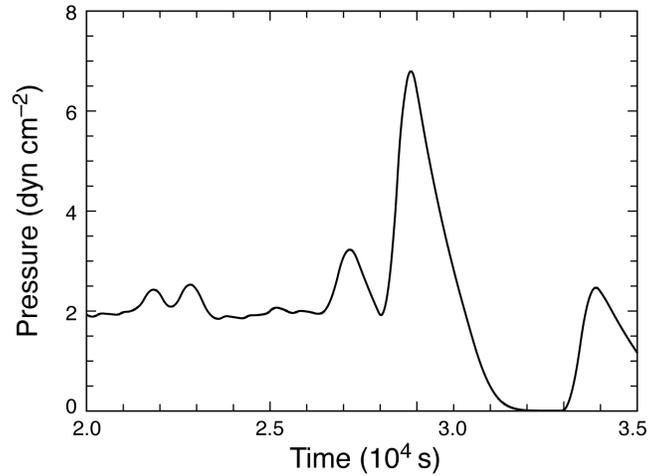

*Figure 5:  Pressure evolution over a 15,000 s interval in the EBTEL simulation (zoom of the dashed curve in Fig. 1)*

## 4.  Coronal Emission

Our objective is to study the basic properties of the coronal emission expected from a simulation of this type. We wish to know whether impulsive heating of the kind that occurs in the simulation is supported or ruled out by observations. Although the predicted emission is only approximate, it should provide guidance as to the feasibility of the basic physical scenario. We choose to emphasize the 193 A channel because there is an abundance of observations in its temperature range.

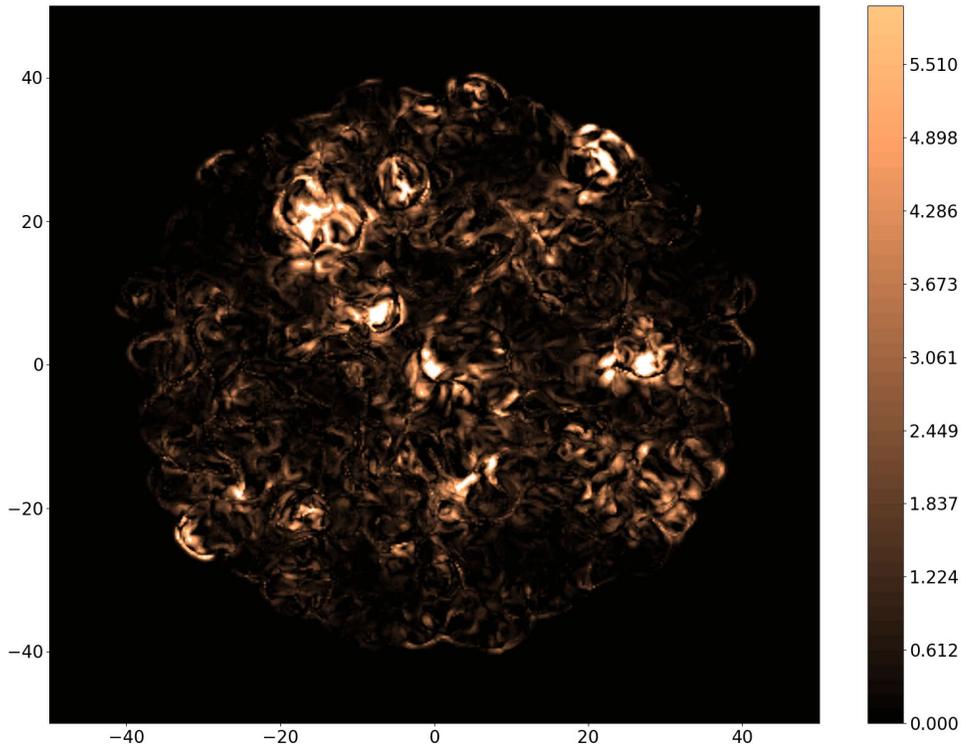

Our model can be used to construct the emissivity throughout the 3D volume, as in the GX_Simulator approach (Nita et al. 2018), but we here choose to concentrate on the emissivity in the midplane. Figure 6 is a map of the 193 A midplane emissivity at $t = 46,797$ s, well after the statistical steady state has been achieved. If our initial field were a magnetic arcade, rather than straight, the midplane would correspond to a vertical cut through the apices of the curved field lines, as shown schematically in Figure 7. The dimensions are 100,000 km x 100,000 km, but this depends on how we choose to convert from code units to physical units, as discussed below.

Figure 8 shows the same map superposed with the projected locations of the vortex cells of the "photospheric" driving. There is no driving at the perimeter and therefore no heating and no emission there. We stress that ubiquitous reconnection events result in a complex magnetic connectivity between the boundary and midplane. The system is not characterized by simple disconnected twisted flux tubes arranged side by side. This is evident in the intricate and continually evolving web of current sheets that is produced by the combination of reconnection and driving (Fig. 4 in Knizhnik et al. 2018). It is also apparent in Figure 9, which shows field lines passing through a regular 5x5 grid of positions in the mid-plane. This is the same time and emissivity map as Figures 6 and 8. The $v_y$ (left-right) component of the boundary flows is shown at the bottom. Note how the field lines are intertwined and map between noncomplementary vortex cells at the top and bottom. Some field lines have widely separated $(x,y)$ positions at the top and bottom.

Figure 6 is also a movie covering the full duration of the simulation. It reveals two basic components to the emission. First, there are many small seemingly uncorrelated brightenings that give the appearance of twinkling throughout the plane. The individual features have a variety of shapes but are often elongated. Their long dimension is generally, but not always, smaller than the driver cell diameter. Second, there are distinct clusters of brightenings that persist for longer than the individual features that comprise them. The clusters have irregular shapes, but with an envelope that is roughly circular in the sense that the aspect ratio is not large.  Each cluster encompasses several driver cells. They, like the individual brightenings – both within and outside of the clusters – have no obvious spatial relationship with the cells. The brightenings do not occur preferentially at cell boundaries or cell centers.

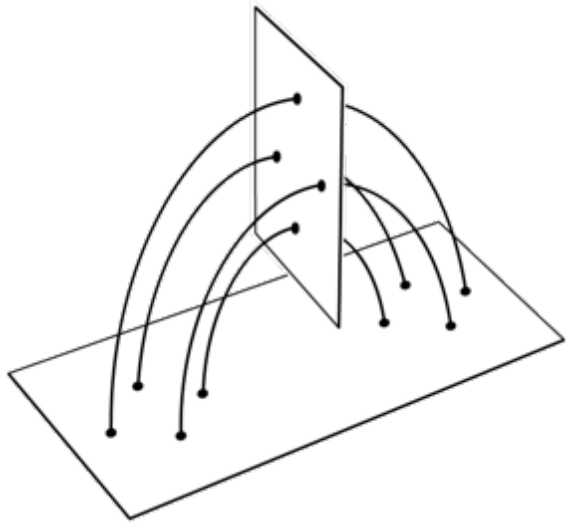

*Figure 7:  Schematic showing how the midplane of the simulation would correspond to a vertical cut if the initial field were an arcade rather than straight.*

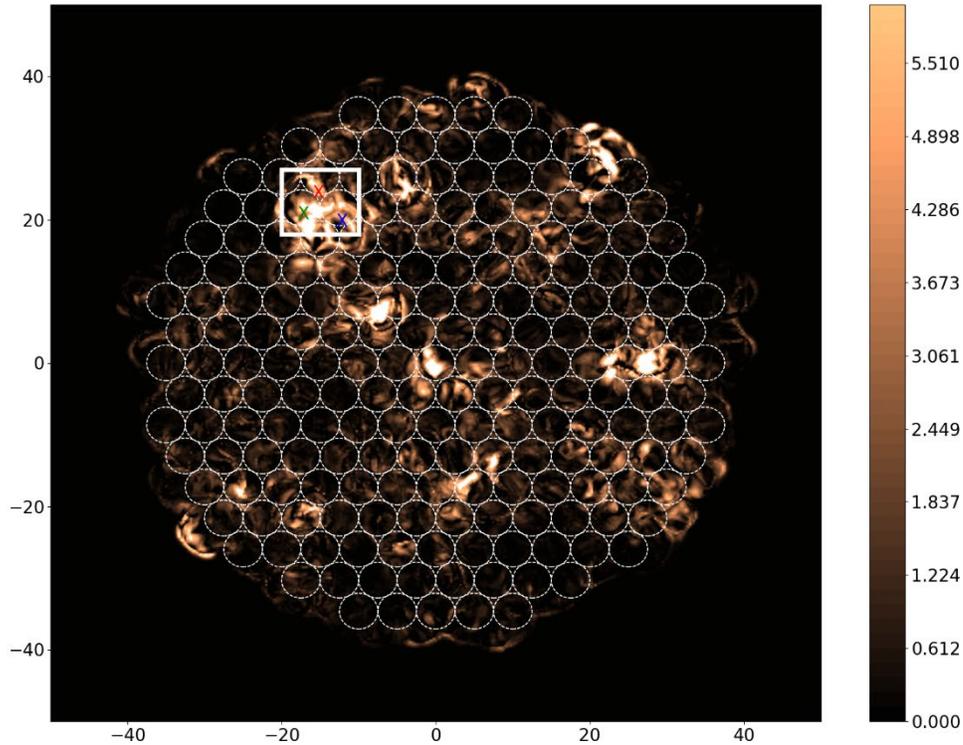

*Figure 8: Map of 193 A emissivity from Fig. 6 overlaid with the projected locations of the vortex driver cells at the "photospheric" boundaries. The cluster of brightenings near position (-17, 22) is shown close up in Fig. 9.*

We stress that the emissivity map is not the same as a coronal image. An image represents a line-of-sight integration of the emissivity through the volume. For example, integrating along a vertical line or horizontal line in the map would give the brightness of a single pixel in an image that corresponds to an observation from above and from the side, respectively, in Figure 7.

Like the emissivity map, coronal images of active regions also have two components: a diffuse component and distinct bright loops. We have suggested previously that the diffuse component is due to random nanoflares, while loops are bundles of spatially unresolved strands that are heated by "storms" of nanoflares (Klimchuk, 2009, 2015). This explanation of loops reconciles several observations that are otherwise difficult to understand. Our combined MHD/cooling simulation is entirely consistent with this observation-based picture. We now explore the agreement in more detail.

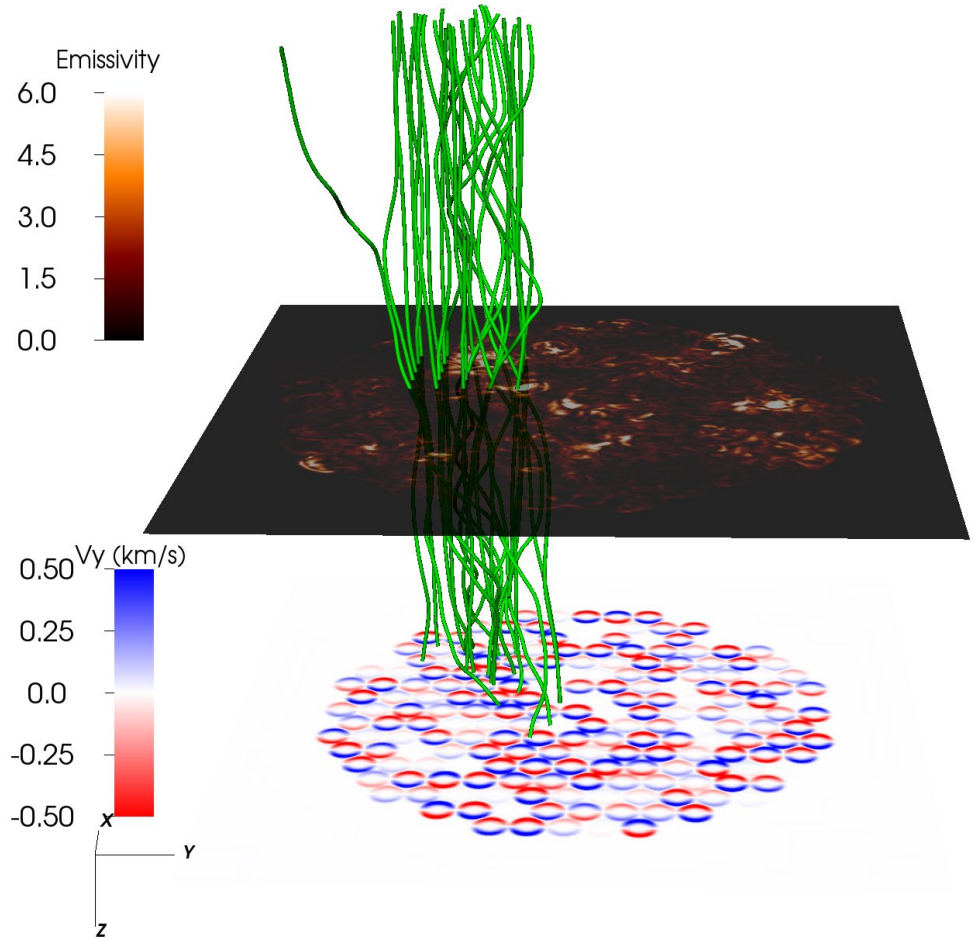

*Figure 9: Magnetic field lines (green), 193 A emissivity in the midplane, and $V_y$ component of driver flow at the lower boundary at t = 46,797 s.*

Figure 10 show a closeup of the cluster near position (-17, 22) in Figure 8, which we equate with a single coronal loop in an image. As already noted, it encompasses several driver cells and has no obvious spatial relationship with the cells. The emission is both highly structured and highly variable, and it is contained within an envelope that is the loop's cross section. The spatial and temporal details are smoothed out by the line-of-sight integration and finite pixel size of a real observation.

Figure 11 shows profiles of intensity versus position for a cut across the loop axis in a hypothetical observation. The black curve corresponds to a view from above and is obtained by integrating along $y$ at $x$ positions in the range $-30 < x < 0$. The red curve corresponds to a view from the side and is obtained by integrating along $x$ at $y$ positions in the range $10 < y < 40$. The integrations are performed over the full dimension of the simulation, not just the close-up region in Figure 10. Thus, the bright feature near position (20, 30) in Figure 8 contributes to the enhanced emission on the right side of the red intensity profile.

These intensity profiles are not unlike those of actual observed loops (Klimchuk & DeForest 2020; Williams et al. 2020). For example, loop emission is generally fainter than background emission in real data. The full width at half maximum (FWHM) of the intensity profiles in Figure 11 is about 7000 km as determined by eye. In comparison, actual loops observed at comparable temperatures have FWHM averaging around 1000 km (Klimchuk 2015). We note that this particular cluster is wider than most clusters in the movie. Also, its size depends on the conversion from dimensionless code units to physical units, which is somewhat arbitrary.

Figure 12 shows light curves for the three locations marked by red, blue, and green X's in Figure 10. There is tremendous variability at all three locations due to the impulsive nature of the heating. Figure 13 shows the light curve for the integrated emission over the white box in Figure 10. The impulsiveness at the individual locations is washed out, and the spatially integrated light curve exhibits a relatively smooth rise and fall. This is precisely the case for real loops. The duration as given by the FWHM is about 10,000 s. In comparison, the observed FWHM of 193 A loops is typically a few thousand seconds, but ranging between 600 and 18,000 s (Winebarger et al. 2003; Ugarte-Urra et al. 2006). Again, the simulation values depend on the assumed conversion to physical units. The spatial and temporal conversions are linked, and if we decrease the spatial conversion by, say, a factor of 5, we must decrease the temporal conversion by the same factor. Doing so would bring both the width and lifetime of the simulation loop into agreement with observed values.

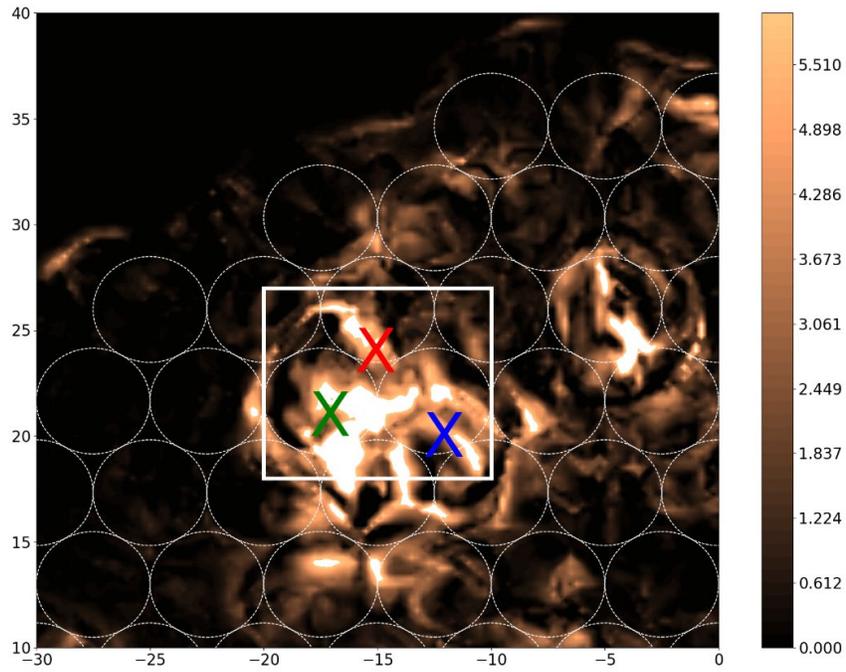

*Figure 10: Close up of Fig. 8 showing a cluster of brightenings. X's indicate the locations of the light curves in Fig. 11. The white box marks the area of the spatially integrated light curve in Fig. 12.*

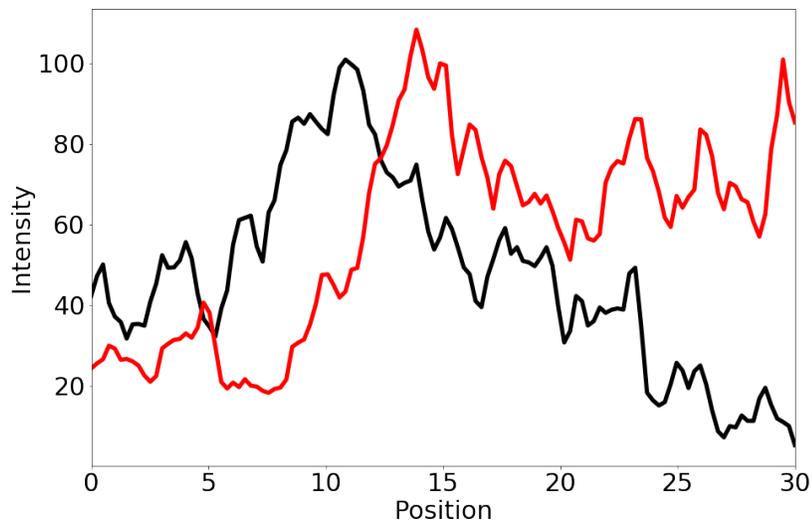

*Figure 11: Intensity profiles (emission integrated along the line of sight versus position) corresponding to an observation made from the top or bottom (black) and from the left or right (red). The spatial coordinates are offset from those in the other figures.*

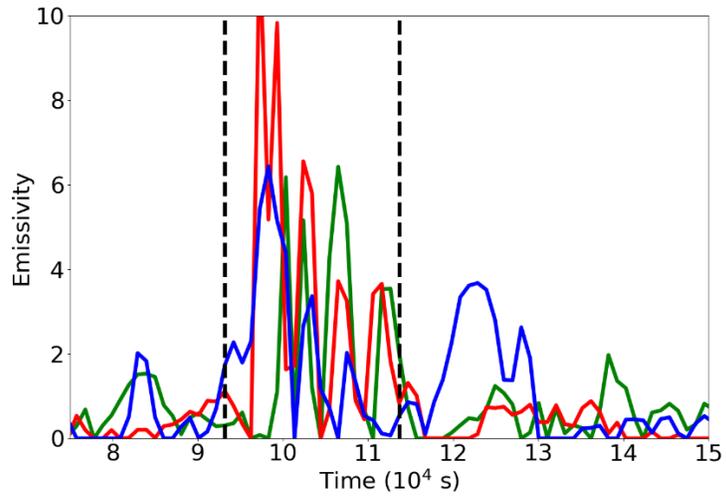

*Figure 12: 193 A light curves (emissivity versus time) at the three locations indicated by the X's in Fig. 9, with corresponding colors. The dashed vertical lines roughly demarcate the lifetime of the complete cluster of brightenings.*

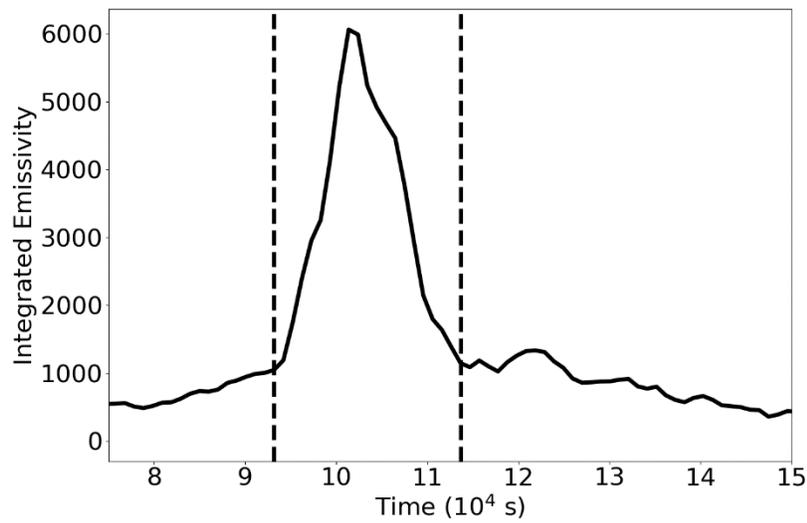

*Figure 13: 193 A light curve for the emissivity integrated over the white box in Fig. 9. The dashed vertical lines roughly demarcate the lifetime of the complete cluster of brightenings.*

Changing the conversion is not without implications. The ratio of the vertical dimension of the simulation box (initial strand length) to the diameter of the driver vortex cells is 4:1. If the cells represent small scale motions associated with photospheric convection, then this ratio is much too small. Reducing the cell diameter by a factor of 5 to 1000 km – so as to better match observed loop widths and lifetimes – results in a very unrealistic loop length of 4000 km. Whether the basic behavior of the heating would be significantly different with more realistic lengths (larger aspect ratios) must be investigated. One potentially important factor is the Alfven travel time along the strand. It is about an order of magnitude larger in our simulation than in real active regions. Another factor is the effect of line tying on the tearing instability that initiates reconnection. The impact of line tying is reduced in longer strands. Note that our choice of a small length to diameter aspect ratio represents a tradeoff between spatial resolution and computational cost.

It is significant that the loop cross section in the emissivity map – the envelope of the cluster – is roughly circular. Recent observational studies find that this is the case for most real loops (Klimchuk & DeForest 2020; Williams et al. 2021; McCarthy et al. 2021). Although this has generally been assumed, there has until now been no physical justification. We have argued that the circular shape in our simulation is not an artifact of the driving, but confirmation with different forms of driving is needed. We agree with Hood et al. (2016) and Reid et al. (2018) that nanoflare storms probably have an avalanche nature, where one event triggers subsequent events. If the process begins at one location and spreads equally in all transverse directions, then a circular shape is expected.

We note the recent interesting study by Malanushenko et al. (2022). They analyzed an MHD simulation of an entire active region and found that many loops in the synthetic images based on the simulation come not from thin tube-like structures, but from large warped veil-like emissivity structures in the 3D volume. Loops are visible in places where the veils are viewed edge on and not where they are viewed face on. We can examine our simulation in that context.

Our idea at the start of this study was to investigate a corona that is driven by small-scale photospheric flows. Our initial normalization based on a strand length of 20,000 km implies driver vortices of 5000 km diameter; however, observations and simulations suggest that a size closer to 1000 km is more appropriate (Bonet et al. 2008; Wedemeyer-Bohm & Rouppe van der Voort 2009; Wedemeyer-Bohm et al. 2012). As noted above, such a normalization brings the diameters and lifetimes of the clusters of brightenings into good agreement with the observed widths and lifetimes of coronal loops. The brightenings themselves have a smaller scale (< 1000 km). They correspond to ribbons in 3D. In comparison, the veils in Malanushenko et al. (2022) tend to be much larger – of order 10,000 km. If we were to normalize our simulation to match this scale, the driver vortices would have a diameter greater than 10,000 km. This may suggest that the flows responsible for the veils in Malanushenko et al. have a scale of this size. We are presently involved in a study to identify the physical origin of the veils.

The above discussion specifically concerns emission in the 193 A channel, but emission observed in the hotter and broader 335 A channel is qualitatively similar. Figure 14 has a snapshot at the same time as the 193 A snapshot and a movie of the full simulation. As with 193 A, there are both random brightenings and clusters of brightenings. The most significant difference is a higher level of quasi-steady, quasi-uniform emission. This is to be expected based on a comparison of the light curves Figures 2 and 3. There are extended periods when relatively high frequency heating maintains the plasma in the temperature range of the 335 A channel and above the range of the 193 A channel. This is consistent with coronal observations. Distinct loops stand out less prominently above the background in 335 A images than in 193 A images.

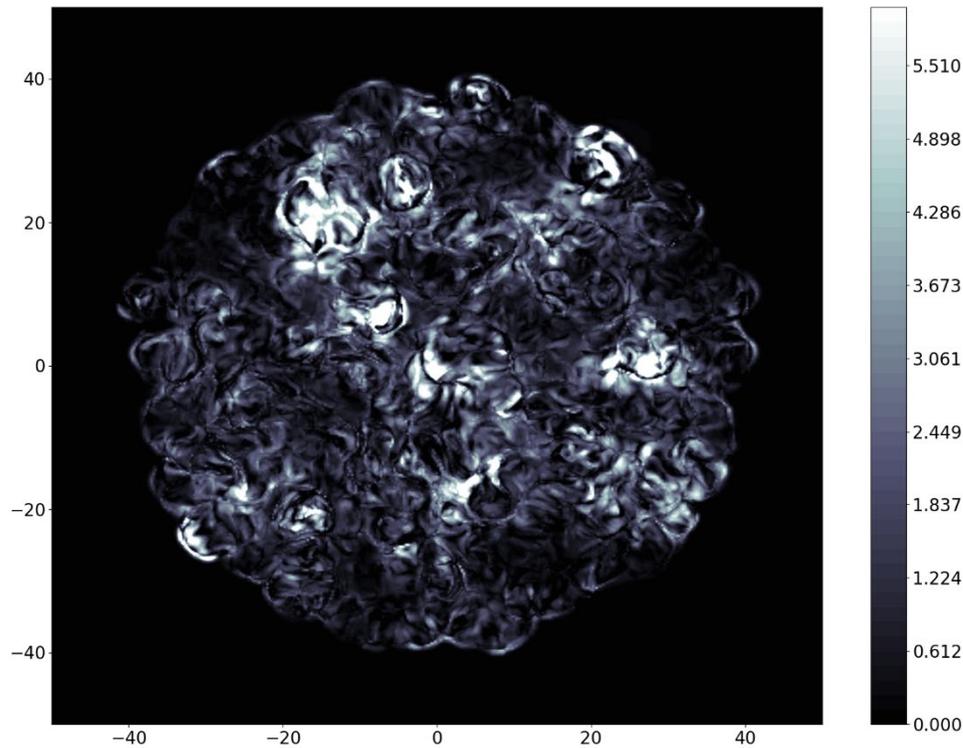

*Figure 14: Map of emissivity as would be detected in the 335 A channel of AIA in the midplane of the simulation at t = 46,797 s. The intensity scale is logarithmic with arbitrary units. An animated version is available covering the full simulation ($2.03 \times 10^5$ s).*

## 5. Plasma Mixing

We return to the question of plasma mixing that occurs when two strands reconnect and exchange sections. This impacts the change in $P_{MHD}$ used to infer coronal heating on the right side of Equation 3. As emphasized in Klimchuk (2015), if the reconnecting strands have different pressures, each new strand will have an average pressure that is intermediate between the two original pressures. The strands equilibrate over the cooling model timestep, as noted above. It is easy to see that the maximum change in pressure from mixing is one-half of the original pressure difference and occurs when the strands reconnect exactly at their midpoints. The pressure change is less if they reconnect closer to the photospheric boundaries. Reconnection very close to the boundaries produces minimal mixing and little pressure change. The average pressure change expected with random reconnection locations is one-quarter of the original pressure difference between the strands.[5]

We estimate the expected magnitude of the mixing effect using our EBTEL simulation discussed in Section 2. If a collection of many similar strands is heated randomly by nanoflares, then the pressures of any two adjacent strands at a given time can be represented by any two random times during the simulation of a single strand. Comparing multiple pairs of random times in our simulation, we find that the magnitude of the pressure difference averages 2.12 dyn cm$^{-2}$. Assuming that the strands reconnect at random locations along their length, we divide by four to estimate a typical pressure change from mixing of 0.53 dyn cm$^{-2}$. This compares to a mean pressure increase from direct heating in the simulation of 0.84 dyn cm$^{-2}$. The effect of mixing is not small overall, but it is negligible compared to larger heating events that produce the brighter emission. As seen in Figure 1, those events have pressure jumps of several dyn cm$^{-2}$.

Unfortunately, the effect of mixing may be exaggerated in the MHD simulation. Radiation is not included, and the uncooled pressures steadily increase over the course of the simulation. This leaves open the possibility that pressure variations across the system may become artificially large. This would cause the mixing during reconnection events to also be artificially large.

We estimate the effect of mixing in the simulation by comparing $P_{MHD}$ at adjacent locations in the mid-plane at a time halfway into the simulation. The solid curve in Figure 15 shows the cumulative probability distribution function for the magnitude of the differences. They have been reduced by a factor of four to give the average mixing under the assumption of random reconnection locations. 75% of the values are less than 0.5 dyn cm$^{-2}$. The dashed curve shows cumulative probabilities for the cooled pressures. Evidently, reconnection and mixing are rather efficient at preventing large pressure variations from developing, even without cooling. Since the pressure increases that occur during the simulation (Fig. 4) are much larger than the expected changes from mixing, we conclude that direct heating dominates, especially in the larger events that contribute the most emission. We also note that the general behavior of the emission does not change over the course of the simulation, as would be expected if mixing began to dominate at later times as $P_{MHD}$ increases.

---

[5] Consider two strands of different pressure and equal length (a good approximation for our simulation, as discussed in Section 5 of Knizhnik et al. 2019 and also evident in Figure 9). Suppose they reconnect a fractional distance $f$ of the strand length from the mid-plane. The average pressure in one new strand increases, while that in the other new strand decreases by the same amount. The unsigned change in pressure is $\Delta P = (1/2 - f) |P_2 - P_1|$. If reconnection occurs at random locations (random $f$ in the range [0, 1/2]), the average unsigned pressure change is $<\Delta P> = 1/4 |P_2 - P_1|$.

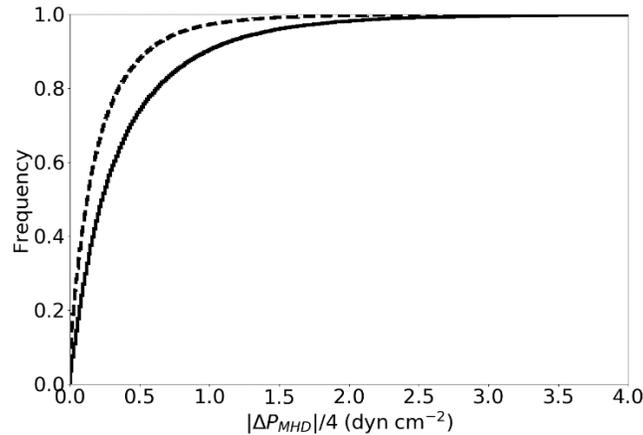

*Figure 15: Cumulative probability distribution function of the pressure difference magnitudes between adjacent grid points in the mid-plane of the MHD simulation, with and without cooling (dashed and solid curves, respectively). Differences have been reduced by a factor of four to approximate the effect of mixing.*

There is one further effect to consider. When we compute the pressure change in a strand to use in the cooling model, Equation 3, we assume that the strand maintains a fixed position in the mid-plane over the 445 s time step, $\Delta t$. If the strand drifts during this time, there will be an artificial change in pressure. We have verified that this effect is very small. Typical horizontal velocities in the mid-plane imply displacements of only a few hundredths of a grid cell dimension over the time step.

## 6. Summary and Future Plans

We have presented a simple method for approximating the coronal emission that could be expected from an MHD simulation that does not include radiation, thermal conduction, or coupling to the lower solar atmosphere. These effects must be accounted for in any meaningful comparison with observations. The method – called the cooling model – is applied to the simulation output *post facto* and thus can be used with any MHD simulation. It could also be incorporated directly into an MHD simulation and operate in real time, but it would not constitute a coupled model since it does not feedback on the MHD. The method operates on the average pressures along field lines. We have applied it to our own previously published simulations of coronal heating that results when an initially straight magnetic field is driven by many small vortex flows at the "photospheric" boundaries. The heating is fundamentally impulsive, taking the form of nanoflares. The results suggest that scattered and uncorrelated nanoflares give rise to the diffuse component of coronal images, while coronal loops are produced by nanoflare storms. Several observed properties of loops are qualitatively reproduced: width, lifetime, and approximately circular shape of the cross-section.

This is not the final word, but it indicates that the simulation has significant merit and gives us encouragement to proceed in this direction. We plan several improvements. We will replace the vortex driver flows with more random forms of driving. This is especially important for verifying or refuting our claim that the quasi-circular nature of the bright clusters (loop cross sections) is not an artifact of the driving. We will consider expanding field geometries rather than the straight uniform field studied so far. This could reveal a nonuniform height distribution of nanoflares, which would have implications for the occurrence of thermal nonequilibrium (e.g., Klimchuk 2019). We will consider longer strands relative the characteristic size of the driver flows to see how the behavior is affected. We will analyze more quantitatively the properties of the nanoflare brightenings, using the same techniques we employed in Knizhnik et al. (2018) to study proxies of heating. We will investigate the causes of the collective behavior that gives rise to nanoflare storms.

Finally, we will replace the simple exponential cooling model with a full EBTEL treatment. EBTEL provides the time dependent coronal temperature and density, allowing a rigorous determination of emissivity. The time dependent heating will still come from the change in pressure in the MHD simulation. Another advantage of EBTEL is that it provides the differential emission measure distribution of the transition region. We can use this to compute the brightness of the transition region in different observing channels. It can rival or exceed the coronal brightness (Schonfeld & Klimchuk 2021; Nita et al. 2018).

This work was supported by the GSFC Internal Scientist Funding Model (competitive work package) program. We thank the referee for comments and suggestions.

## References

Antolin, P., Okamoto, T. J., De Pontieu, B., Uitenbroek, H., Van Doorsselaere, T., et al. 2015, ApJ, 809, 72

Barnes, W. T., Bradshaw, S. J., & Viall, N. M. 2021, ApJ, 919, 132

Bonet, J. A., Marquez, I., Sanchez Almeida, J., Cabello, I., & Domingo, V. 2008, ApJ, 687, L131

Cargill, P. J. 1993, Sol. Phys., 147, 263

Cargill, P. J. 1994, ApJ, 422, 381

Cargill, P. J., & Bradshaw, S. J. 2013, ApJ, 772, 40

Cargill, P. J., Bradshaw, S. J., & Klimchuk, J. A. 2012, ApJ, 752, 161

DeVore, C. R., & Antiochos, S. K. 2008, ApJ, 680, 740

Hood, A. W., Cargill, P. J., Browning, P. K., & Tam, K. V. 2016, ApJ, 817, 5

Klimchuk, J. A. 2006, Solar Phys., 234, 41

Klimchuk, J. A. 2009, in The Second Hinode Science Meeting: Beyond Discovery—Toward Understanding (ASP Conf. Ser. Vol. 415), ed. B. Lites, M. Cheung, T. Magara, J. Mariska, & K. Reeves (San Francisco: Astron. Soc. Pacific), p. 221

Klimchuk, J. A. 2015, Phil. Trans. R. Soc. A, 373, 20140256

Klimchuk, J. A. 2019, Solar Phys., 294, 173

Klimchuk, J. A., & DeForest, C. E. 2020, ApJ, 900, 167

Klimchuk, J. A. and Hinode Review Team 2019, PASJ, 71 (5), R1 (doi: 10.1093/pasj/psz084)

Knizhnik, K. J., Antiochos, S. K., & DeVore, C. R. 2015, ApJ, 809, 137

Knizhnik, K. J., Antiochos, S. K., & DeVore, C. R. 2017a, ApJ, 835, 85

Knizhnik, K. J., Antiochos, S. K., DeVore, C. R., & Wyper, P. F. 2017b, ApJL, 851, L17

Knizhnik, K. J., Antiochos, S. K., Klimchuk, J. A., & DeVore, C. R. 2019, ApJ, 883, 26

Knizhnik, K. J., Barnes, W. T., Reep, J. W., & Uritsky, V. M. 2020, ApJ, 899, 156

Knizhnik, K.J., & Reep, J.W. 2020, Solar Phys., 295, 21

Knizhnik, K. J., Barnes, W. T., Reep, J. W., & Uritsky, V. M. 2020, ApJ, 899, 156

Knizhnik, K. J., Uritsky, V. M., Klimchuk, J. A., & DeVore, C. R. 2018, ApJ, 853, 82

Leake, J. E., Daldorff, L. K. S., & Klimchuk, J. A. 2020, ApJ, 891, 62

Lemen, J. R., Title, A. M., Akin, D. J., et al. 2012, Solar Phys., 275, 17

Lopez Fuentes, M., & Klimchuk, J. A. 2016, ApJ, 828, 86

Malanushenko, A., Cheung, M. C. M., DeForest, C. E., Klimchuk, J. A., & Rempel, M. 2022, ApJ, 927, 1

McCarthy, M. I., Longcope, D. W., & Malanushenko, A. 2021, ApJ, 913, 56

Nita, G. M., Viall, N. M., Klimchuk, J. A., Loukitcheva, M. A., Gary, D. E. et al. 2018, ApJ, 853, 66

Reale, F. 2014, LRSP, 11, 4

Reale, R., & Landi, E. 2012, A&A, 543, A90

Reid, J., Hood, A. W., Parnell, C. E., Browning, P. K., & Cargill, P. J. 2018, A&A, 15, A84

Ugarte-Urra, I., Winebarger, A. R., & Warren, H. P. 2006, ApJ, 643, 1245

Van Ballegooijen, A. A., Asgari-Targhi, M., Cranmer, S. R., & DeLuca, E. E. 2011, ApJ, 736, 3

Viall, N. M. & Klimchuk, J. A. 2011, ApJ, 738, 24

Wedemeyer-Bohm, S., & Rouppe van der Voort, L. 2009, A&A, 507, L9

Wedemeyer-Bohm, S., et al. 2012, Nature, 486, 505

Williams, T., Walsh, R. W., & Morgan, H. 2021, ApJ, 919, 47

Winebarger, A. R., Warren, H. P., & Seaton, D. B. 2003, ApJ, 593, 1164